\documentstyle[12pt,epsf,twoside,espcrc1]{article}

\newcommand{\AmS}{{\protect\the\textfont2
  A\kern-.1667em\lower.5ex\hbox{M}\kern-.125emS}}

\newcommand{\beq}{\begin{equation}}
\newcommand{\eeq}[1]{\label{#1} \end{equation}}
\newcommand{\beqar}{\begin{eqnarray}}
\newcommand{\eeqar}[1]{\label{#1} \end{eqnarray}}
\newcommand{\insertplot}[1]{ \begin{center}\leavevmode\epsfysize=
7.5cm
\epsfbox{#1}\end{center}}

\title{Freeze-out in hydrodynamical models in relativistic heavy 
ion collisions}

\author{
V.K. Magas\address{Section for Theoretical Physics, 
Department of Physics\\
University of Bergen, Allegaten 55, 5007 Bergen, Norway}, 
Cs. Anderlik\address{Section for Theoretical Physics, 
Department of Physics\\
University of Bergen, Allegaten 55, 5007 Bergen, Norway},  
L.P. Csernai\address{Section for Theoretical Physics, 
Department of Physics\\
University of Bergen, Allegaten 55, 5007 Bergen, Norway\\
KFKI Research Institute for Particle and Nuclear Physics\\
P.O.Box 49,  1525 Budapest, Hungary},  
F. Grassi\address{Instituto de F\'{\i}sica, Universidade de 
Sao Paulo\\
CP 66318, 05389-970 S\~ao Paulo-SP, Brazil},
W. Greiner\address{Institut f\"ur Theoretische Physik, 
Universit\"at Frankfurt\\
Robert-Mayer-Str. 8-10, D-60054 Frankfurt am Main, Germany},
Y.~Hama\address{Instituto de F\'{\i}sica, Universidade de Sao Paulo\\
CP 66318, 05389-970 S\~ao Paulo-SP, Brazil},
T.~Kodama\address{Instituto de F\'{\i}sica, Universidade Federal do 
Rio de Janeiro\\
CP 68528, 21945-970 Rio de Janeiro-RJ, Brazil},
Zs.I. L\'az\'ar\address{Section for Theoretical Physics, 
Department of Physics\\
University of Bergen, Allegaten 55, 5007 Bergen, Norway} 
and
H. St\"ocker\address{Institut f\"ur Theoretische Physik, 
Universit\"at Frankfurt\\
Robert-Mayer-Str. 8-10, D-60054 Frankfurt am Main, Germany}}

\begin{document}
\maketitle

\section{INTRODUCTION}
In continuum and
fluid dynamical models, 
particles, which leave the system and reach the detectors, can be
taken into account via freeze-out
(FO) or final break-up schemes, where the frozen out particles are 
formed on
a \mbox{3-dimensional} hypersurface in space-time. Such FO
descriptions are
important ingredients of evaluations of two-particle correlation data, 
\mbox{transverse-,} \mbox{longitudinal-,} radial- and cylindrical-
flow analyses,
transverse momentum and transverse mass spectra and many other 
observables.
The FO on a hypersurface is a discontinuity, where the pre FO 
equilibrated and
interacting matter abruptly changes to non-interacting particles, 
showing an
ideal gas type of behavior.

The frequently used Cooper-Frye formula,
the rapidity distribution,
transverse momentum spectrai, etc., -- all include expression 
$f_{FO}(x,p,T,n,u^\nu) 
\ p^\mu d\hat{\sigma}_\mu\,,$
 where $f_{FO}(x,p,T,n,u^\nu)$ is the post FO distribution
which is unknown from the fluid dynamical
model, $d\hat{\sigma}_\mu$ is normal vector to the FO hypersurface.
Those formulas work well for timelike $d\hat{\sigma}^\mu$ 
($p^\mu d\hat
{\sigma}_\mu > 0$). If $d\hat{\sigma}_\mu$ is spacelike we count 
particles 
going backwards through FO front as well as outwards. The post FO
distribution can not be a thermal one! In fact $f_{FO}$ should contain 
only particles
which cross the FO front outwards, $p^\mu d\hat
{\sigma}_\mu > 0$, since the rescattering and back scattering are not 
allowed any more in the post FO side.
\beq
f_{FO}(x,p,T,n,u^\nu,d\hat{\sigma}_\mu)=0, \ \ \ \ p^\mu d\hat
{\sigma}_\mu < 0\,.
\eeq{Int3}

Initially if we know the pre FO baryon current and energy-momentum 
tensor,
$N_0^\mu$ and $T_0^{\mu\nu}\ ,$ 
we can calculate locally, across a surface element of normal vector
$d\hat{\sigma}^\mu$
the post FO quantities, 
$N^\mu$ and $T^{\mu\nu}$, 
from the relations:
$
[N^\mu\ d\hat{\sigma}_\mu] = 0 
$
and
$
[T^{\mu\nu}\ d\hat{\sigma}_\mu] = 0 ,
$
where $[A]\equiv A - A_0$.
One should also check that entropy is nondecreasing in FO.
In numerical calculations the local FO surface can be determined 
most accurately via self-consistent iteration.
Initial ideas to improve the Cooper-Fry FO description this way  
were suggested in refs. \cite{Bu96,CLM97,NL97}.

\section{FREEZE-OUT DISTRIBUTION FROM KINETIC THEORY}

Let us assume an infinitely long tube
with its left half ($x<0$) filled with nuclear matter and in the 
right 
vacuum is maintained. We can remove the dividing wall at $t=0$, and 
then 
the matter will expand into the vacuum. Continuously removing
particles at the right end of the tube and supplying particles on the 
left end we can establish a stationary flow in the tube, where the
particles will gradually freeze out in an exponential rarefraction 
wave
propagating to the left in the matter.  We can move with this front, 
so
that we describe it from the reference frame of the front (RFF).

We assume that there are two components of our momentum 
distribution:   
$f_{free}(x,\vec{p})$  and
$f_{int}(x,\vec{p})$. However, we assume that at $x=0$, $f_{free}$
vanishes exactly and $f_{int}$ is an ideal J\"uttner distribution,
then $f_{int}$ gradually
disappears and $f_{free}$ gradually builds up, according to the
 differential equations:
\beqar
\partial_x f_{int}(x,\vec{p})   dx = - \Theta(p^\mu d\hat{\sigma}_\mu) 
                                   \frac{\cos \theta_{\vec{p}} }
{\lambda}
           f_{int}(x,\vec{p})   dx \,,
\nonumber \\ & & \\
\partial_x f_{free}(x,\vec{p})  dx = + \Theta(p^\mu d\hat{\sigma}_\mu) 
                                   \frac{\cos \theta_{\vec{p}} }
{\lambda}
           f_{int}(x,\vec{p})   dx \,, \nonumber
\eeqar{first}
where $\cos \theta_{\vec{p}}=\frac{p^x}{p}$ in RFF. It expresses the 
fact that particles with momenta orthogonal to the FO surface 
leave the system with bigger probability than particles emitted at an 
angle.

Such a dramatically oversimplified model can reproduce 
cut J\"uttner distribution as a post FO one, but it does not allow the 
complete FO  --  the interacting component of momentum
distribution survives  even if
$x \longrightarrow \infty$.  

To improve our model we take into account rescattering 
within the interacting  component, which  will lead to
re\--ther\-ma\-li\-za\-tion 
and re-equilibration of this component.
We use the  relaxation 
time approximation to simplify the description of the dynamics.
Thus, 
the two components of the momentum distribution develop according to 
the
differential equations:
\beq
\begin{array}{rll}
\partial_x f_{int}(x,\vec{p}) dx =& - \Theta(p^\mu d\hat{\sigma}_\mu) 
                                   \frac{\cos \theta_{\vec{p}} }
{\lambda}
           f_{int}(x,\vec{p})   dx+
 \\ & & \\
                                  
           &+\left[ f_{eq}(x,\vec{p}) -  f_{int}(x,\vec{p})\right]
           \frac{1}{\lambda'} dx, 
\end{array}
\eeq{kin-2}
\beq
\begin{array}{rll}
\partial_x f_{free}(x,\vec{p}) dx =& + \Theta(p^\mu d\hat{\sigma}_\mu) 
                                   \frac{\cos \theta_{\vec{p}} }
{\lambda}
           f_{int}(x,\vec{p})  dx.
\end{array}
\eeq{kin-3}
The interacting component of the momentum distribution, described by 
eq. (\ref{kin-2}), shows the tendency to approach an equilibrated
distribution with a relaxation length  $\lambda' $.  Of course,
due to the energy, momentum and  particle drain, this distribution
$f_{eq}(x,\vec{p})$ is not the same as the initial J\"uttner 
distribution,
but its parameters, $n_{eq}(x)$, $T_{eq}(x)$ and $u^\mu_{eq}(x)$,
change as required by the conservation laws.

Let us assume 
that $\lambda' \ll \lambda$, i.e. re-thermalization is much faster
than particles are freezing out, or much faster than parameters,
$n_{eq}(x)$, $T_{eq}(x)$ and $u^\mu_{eq}(x)$ change.
Then 
$
f_{int}(x,\vec{p})\approx f_{eq}(x,\vec{p})
$,
for 
$
\lambda'\ll \lambda\,.
$
For $f_{eq}(x,\vec{p})$ we assume the spherical J\"uttner form at
any $x$ including both positive and negative momentum parts 
with parameters $n(x),\ T(x)$ and $u_{RFG}^\mu(x)$.
(Here $u_{RFG}^\mu(x)$ is the actual flow velocity of the
interacting, J\"uttner component, i.e. the velocity of the Rest Frame 
of the
Gas (RFG) \cite{Bu96}).
In this case the changes of conserved quantities due to particle drain 
or 
transfer can be  evaluated for an infinitesimal $dx$ and 
new parameters of $f_{eq}(x+dx,\vec{p})$ can be found \cite{ALC98}. 

We would like to point out that, although 
for the spherical J\"uttner distribution  
the Landau and Eckart flow velocities are the same,
the change of this flow velocity  calculated from the loss of baryon 
current
and from the loss of energy current appear to be different
$
du^\mu_{i,E,RFG}(x) \ne  du^\mu_{i,L,RFG}(x) \,.
$
This is a clear consequence of the asymmetry caused by the FO
process as it was discussed in ref. \cite{ALC98,AC98}.
This problem does not occur for the freeze-out of baryonfree 
plasma. 


\begin{figure}[htb]
        \insertplot{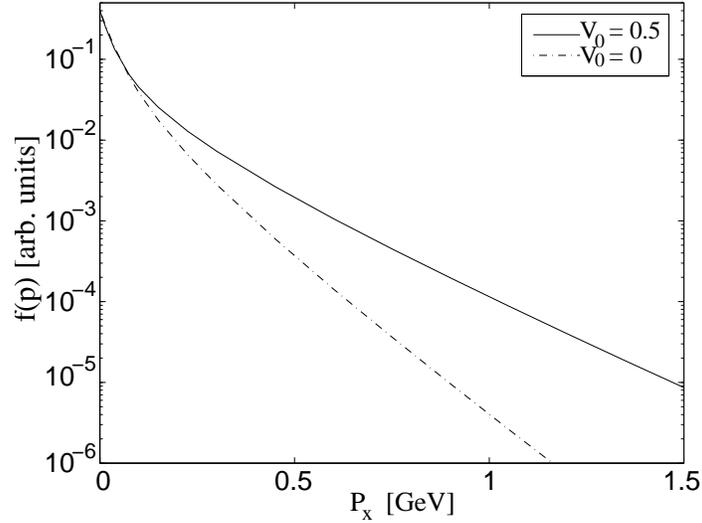}
\caption{The 
local transverse momentum (here $p_x$)
distribution for baryonfree, massless gas 
at $p_y = 0$, $x=100 \lambda$ and $T_0 = 130\,$MeV.  
The transverse momentum spectrum is obviously curved
due to the freeze-out process, particularly for large
initial flow velocities. The apparent slope parameter
increases with increasing transverse momentum. This behavior 
agrees
with observed pion transverse mass spectra at SPS \cite{na44}. 
From \cite{9ath99}.
}

\label{fig:1}
\end{figure}

We performed the calculations, according to this model, for the 
baryonfree 
and massless gas \cite{9ath99,PL99}.
We would like to note that now $f_{int}(x,\vec{p})$ does not
tend to the cut J\"uttner distribution in the limit $x \rightarrow 
\infty$. 
Furthermore, we obtain that $T \rightarrow 0$,
when $x \rightarrow \infty$. 
So, $f_{int}(x,\vec{p})=\frac{1}{(2\pi\hbar)^3}\exp[(\mu-p^{\nu}
u_{\nu})/T] \rightarrow 0$, 
when $x \rightarrow \infty$. Thus, all particles freeze
out in the improved model, but such a physical FO requires 
infinite distance (or time). This second problem may also be 
removed by using volume emission model 
discussed in \cite{9ath99}. 
The application of our onedimensional model to transverse expansion 
gives 
result showing in Fig. 1, which is in qualitative agreement with 
experiment. 
  
 Recent calculations have confirmed that the FO hypersurface
 idealization can be justified even in microscopic reaction models 
(like
 UrQMD or QGSM) for nucleon data in collisions of massive heavy ions.

 The improvements presented here are essential and lead to
 non-negligible qualitative and quantitative changes in
 calculations including FO. Several further details and consequences
 of this improved approach have to be worked out
 still (e.g. \cite{BG99}) to obtain more accurate data from the 
 numerous
 continuum and fluid dynamical models used for the
 description of heavy ion reactions.

\vfill\eject
\end{document}